\documentclass{fundam}
\usepackage{url}
\usepackage{amsmath,amsfonts}
\usepackage{amssymb}
\usepackage{dsfont}
\usepackage{graphicx}
\usepackage{float}
\usepackage{subfig}
\usepackage{color}
\usepackage[a4paper, total={6in, 8in}]{geometry}

\usepackage{tabularx, booktabs}
\newcolumntype{Y}{>{\centering\arraybackslash}X}

\DeclareMathOperator{\Radon}{\mathcal{R}}

\newcommand{\ra}[1]{\renewcommand{\arraystretch}{#1}}

\begin{document}
\setcounter{page}{1001}
\issue{XXI~(2001)}

\title{Improving analytical tomographic reconstructions through consistency conditions}

\address{filippo.arcadu@psi.ch}

\author{Filippo Arcadu\corresponding\\
Institute for Biomedical Engineering, ETH Zurich, 8092 Zurich, Switzerland\\
Swiss Light Source, Paul Scherrer Institute, 5232 Villigen, Switzerland\\
filippo.arcadu{@}psi.ch
\and Jakob Vogel\\
Swiss Light Source, Paul Scherrer Institute, 5232 Villigen, Switzerland\\
federica.marone{@}psi.ch
\and Marco Stampanoni\\
Institute for Biomedical Engineering, ETH Zurich, 8092 Zurich, Switzerland\\
Swiss Light Source, Paul Scherrer Institute, 5232 Villigen, Switzerland\\
stampanoni{@}biomed.ee.ethz.ch
\and Federica Marone\\
Swiss Light Source, Paul Scherrer Institute, 5232 Villigen, Switzerland\\
federica.marone{@}psi.ch} \maketitle


\begin{abstract}
This work introduces and characterizes a fast parameterless filter based on the Helgason-Ludwig
consistency conditions, used to improve the accuracy of analytical reconstructions of
tomographic undersampled datasets. The filter, acting in the Radon domain, extrapolates
intermediate projections between those existing. The resulting sinogram, doubled in views,
is then reconstructed by a standard analytical method.
Experiments with simulated data prove that the peak-signal-to-noise ratio of the results computed by filtered
backprojection is improved up to 5--6 dB, if the filter is used prior to reconstruction.
\end{abstract}


\section{Introduction}
The Radon transform \cite{Radon1917} of a function belonging to the space of rapidly decreasing $C^{\infty}$ functions
on $\mathbb{R}^{2}$ satisfies the Helgason-Ludwig consistency conditions \cite{Ludwig2010,Helgason1981} (HLCC). These 
properties characterize any $k$-th moment of the Radon transform. The consistency conditions, known since
the 1960s \cite{Ludwig2010}, have been mainly exploited in iterative algorithms to restore limited-angle tomographic
datasets \cite{Prince1990,Kudo1991,VanGompel2006,Xu2010}, to reduce translational motion artifacts
in fan-beam computed tomography (CT) reconstructions \cite{Yu2007} and for alignment in cardiac position emission tomography (PET) and CT
\cite{Alessio2010}.
\newline
The algorithm presented in this work utilizes the HLCC for a different purpose, namely, augmenting the number of views
of a sinogram with projections homogeneously acquired in $[0,\pi)$ to improve its reconstruction with an analytical
technique like filtered backprojection \cite{Herman1979} (FBP).
This augmentation strategy should provide better analytical reconstructions especially for undersampled
datasets, i.e., $m \ll n\pi/2$, where $m$ is the number of views and $n$ the number of detector pixels \cite{Kak2001}.
Iterative algorithms, often utilized for the reconstruction of strongly undersampled datasets, are characterized by a
high computational cost, a rather large hyper-parameter space and need for sample-specific constraints and/or a priori knowledge.
The proposed algorithm, instead, acting as a filter that preprocesses the sinogram before
the actual reconstruction, is parameterless, sample-independent and fast (only slightly impacting the total reconstruction speed).
\newline
The use of this HLCC-based filter substantially improves the quality of FBP reconstructions of undersampled datasets. 
However, the reconstruction quality provided by highly optimized iterative algorithms, 
set with finely tuned hyper-parameters and sample-specific a priori knowledge, is not yet achieved. 
Nonetheless, the proposed method represents an effective strategy 
when sufficient computational power (e.g. access to high performance computing facility) is 
not available or high variability of the samples requires time consuming
integration of a priori knowledge and hyper-parameters optimization for each single investigated object.


\section{Helgason-Ludwig consistency conditions}
\label{lhcc}
\begin{figure}[h!]
    \centering
    \hspace*{0.0cm}\includegraphics[width=2.2in]{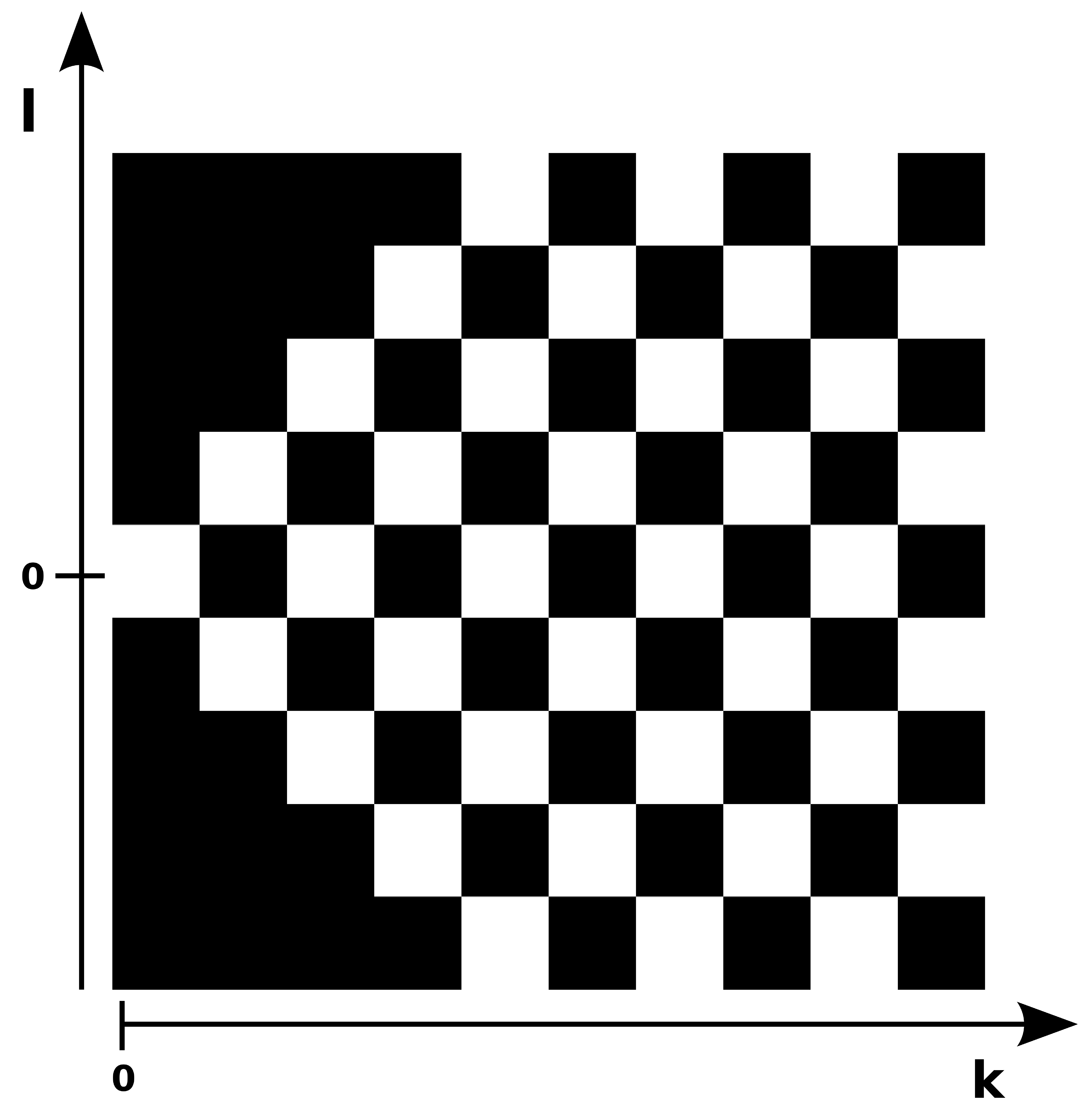}
    \caption{Checkboarded wedge pattern of the $b_{kl}$ coefficients in the $(k,l)$-space. The non-zero $b_{kl}$'s are placed in the white
             squares, the rest is zero.}
    \label{checkboarded-wedge}
\end{figure}

The HLCC are properties characterizing the Radon transform, $\Radon$, of a 2D function
$f(\mathbf{x})$ with bounded support, $\Omega$, entirely placed within the field-of-view (FOV),
i.e., $\Omega \subset $ FOV. In parallel beam geometry, the Radon transform is defined as \cite{Kak2001}:
\begin{equation}
  p(\theta,t) = \Radon\{f(\mathbf{x})\}(\theta,t) = \int\limits_{\Omega}\:d\mathbf{x}\:
                                                   f(\mathbf{x})\:\delta(\mathbf{x}\cdot\hat{\mathbf{n}} - t) \hspace{0.3cm},
  \label{radon-transform}
\end{equation}
where $\mathbf{x} = (x_{1},x_{2})$, $\hat{\mathbf{n}} = (\cos\theta,\sin\theta)$, $\theta$ is the projection angle
formed with the $x_{1}$-axis and $t$ is the distance of the X-ray line from the 
origin of the reference frame.
\newline
The HLCC state that the integral:
\begin{equation}
 \int\limits_{-1}^{+1}dt\:t^{k}\:p(\theta,t)
 \label{lhcc-1}
\end{equation}
is a homogeneous polynomial of degree $k$ in $\cos\theta$ and $\sin\theta$ for $k \geq 0$
\cite{Ludwig2010,Helgason1981}.
The zeroth-order condition ($k = 0$), e.g, corresponds to \cite{Prince1990}:
\begin{equation}
  \mu = \int\limits_{-1}^{+1}dt\:p(\theta,t) = \int\limits_{\mathbb{R}^{2}} d\mathbf{x}\: f(\mathbf{x})
  \hspace*{0.4cm}\forall \:\theta\,,
\end{equation}
meaning that the integral of any projection along $t$ is constant and equals the integral of
$f$ over $\mathbb{R}^{2}$. For $k\geq1$, conditions related to higher moments of $p(\theta,t)$ 
are obtained.
\newline
An effective way to enforce the HLCC is by expanding $p(\theta,t)$ onto a Fourier-Chebyshev basis \cite{Kudo1991}:
\begin{equation}
  p(\theta,t) = \frac{1}{\pi} \sum\limits_{k=0}^{\infty}\:\sum\limits_{l=-\infty}^{+\infty}
                b_{kl}\:(1-t^{2})^{1/2}\:U_{k}(t)\:e^{il\theta} \hspace{0.3cm},
  \label{fourier-chebyshev}
\end{equation}
where $U_{k}(t)$ represents the $k$-th order Chebyshev polynomial of second kind,
the $b_{kl}$'s are the Fourier Chebyshev coefficients and $\theta \in [0,2\pi]$ and
$i$ is the imaginary unit.
On this orthogonal basis, the HLCC become \cite{Kudo1991}:
\begin{equation}
  b_{kl} = 0 \hspace*{0.4cm}\text{for}\hspace{0.3cm} \begin{cases}
							|l| > k  \\
							k + |l| = 2z + 1 & \forall \hspace{0.1cm} z \in \mathbb{N}
                                                   \end{cases}\hspace{0.3cm}.
  \label{bkm-cond}
\end{equation}
If the Radon transform is sampled over $2\pi$,
the positions of the non-zero $b_{kl}$ form in the $(k,l)$-space a checkboarded 
wedge as shown in Fig. \ref{checkboarded-wedge}.
\newline
Consistency conditions have been extended to fan-beam \cite{Chen2005} and cone-beam \cite{Clackdoyle2013} geometry as well.


\section{Proposed method}
\label{proposed-method}
\begin{figure}[!t]
  \centering
  \hspace*{0.0cm}\subfloat[]{\includegraphics[width=2.5in]{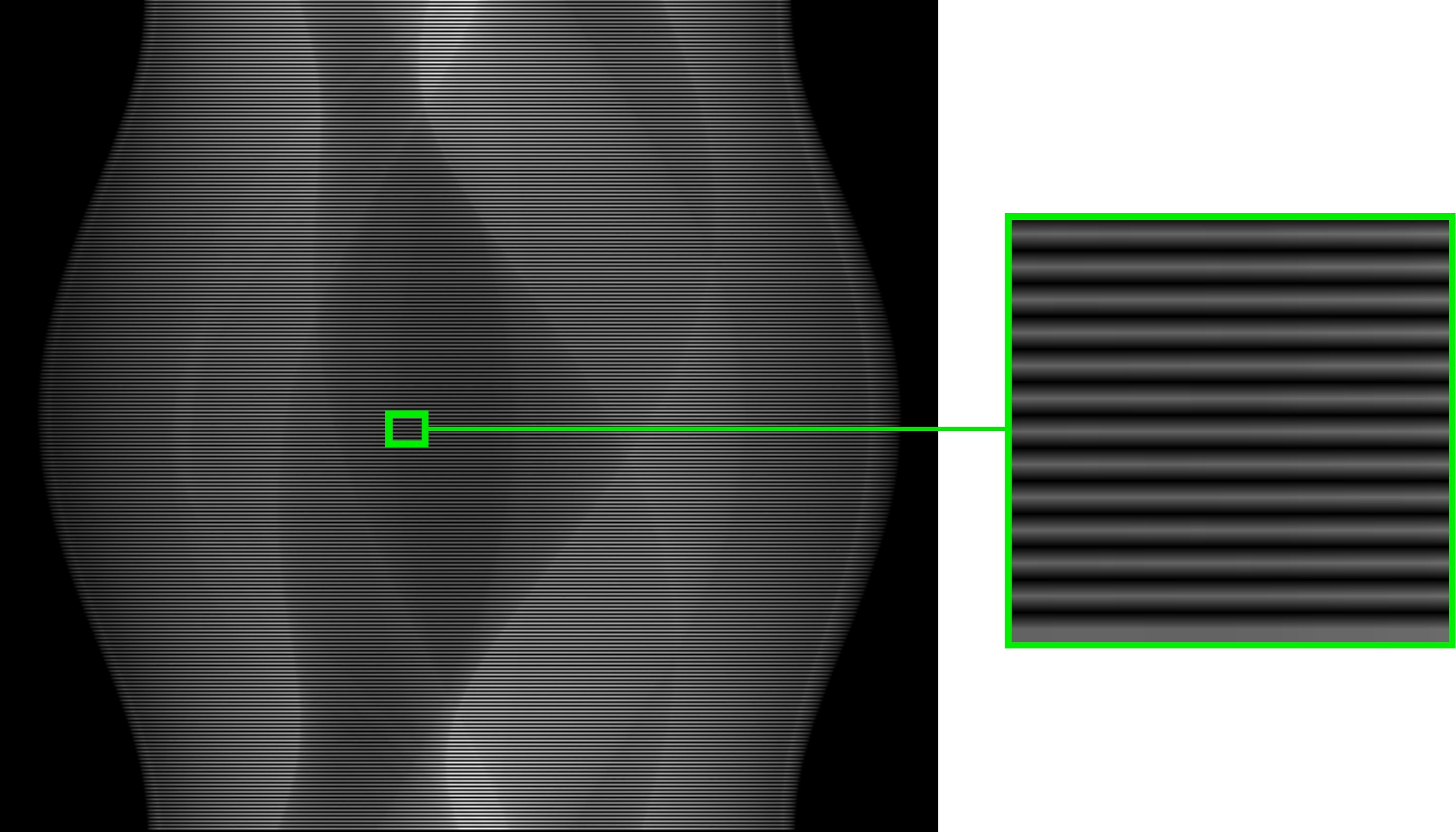}\label{fig:sub1}}%
  \hspace*{2.0cm}\subfloat[]{\includegraphics[width=1.55in]{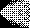}\label{fig:sub2}}
  \caption{Example of a sinogram interleaved with 0-valued lines (a) and the binarized real part of its
           Fourier-Chebyshev decomposition (b), showing that inconsistencies give rise to non-zero $b_{kl}$'s
           at the locations indicated in (\ref{bkm-cond}).}
  \label{proof-of-principle}
\end{figure}

The discrete version of $p(\theta,t)$ (also called sinogram) is indicated with $p[\theta_{h},t_{j}]$, where
$\theta_{h} = h\pi/m \in [0,\pi)$ for $h = 0,1, ...,m-1$ and $t_{j} = -1 + 2j/(n-1)$ for $j = 0, 1,...,n-1$,
considering $t_{j} \in [-1,1]$.
\newline
The proposed Helgason-Ludwig sinogram filter (HLSF) doubles the number of views
of a sinogram $\in \mathbb{R}^{m\times n}$ by extrapolating $m$ missing projections at
intermediate angles $\theta_{h+1/2} = (\theta_{h} + \theta_{h+1})/2$.
The input sinogram is, first, interleaved with 0-valued projections, as shown in Fig. \ref{proof-of-principle}(a).
In this way, it becomes an inconsistent dataset
characterized by several non-zero $b_{kl}$'s at the locations indicated in (\ref{bkm-cond}) (Fig. \ref{proof-of-principle}(b)).
By enforcing (\ref{bkm-cond}),
consistency is recovered, the 0-valued lines are filled with extrapolated values and 
the resulting sinogram $\in \mathbb{R}^{2m\times n}$ can be reconstructed by means of an analytical method like FBP.
\newline
HLSF consists of the following four steps:
\begin{enumerate}
 \item extension of the data to $[0,2\pi]$;
 \item creation of an intermediate sinogram with $4m$ views; the $2m$ original
       projections are interleaved with $2m$ zero lines;
 \item imposition of the HLCC on the intermediate inconsistent sinogram;
 \item crop of the interval $[0,\pi)$ and reassignment of the original $m$ views.
\end{enumerate}

\begin{figure}[!t]
  \centering
  \hspace*{-1.3cm}\subfloat[]{\includegraphics[width=1.2in]{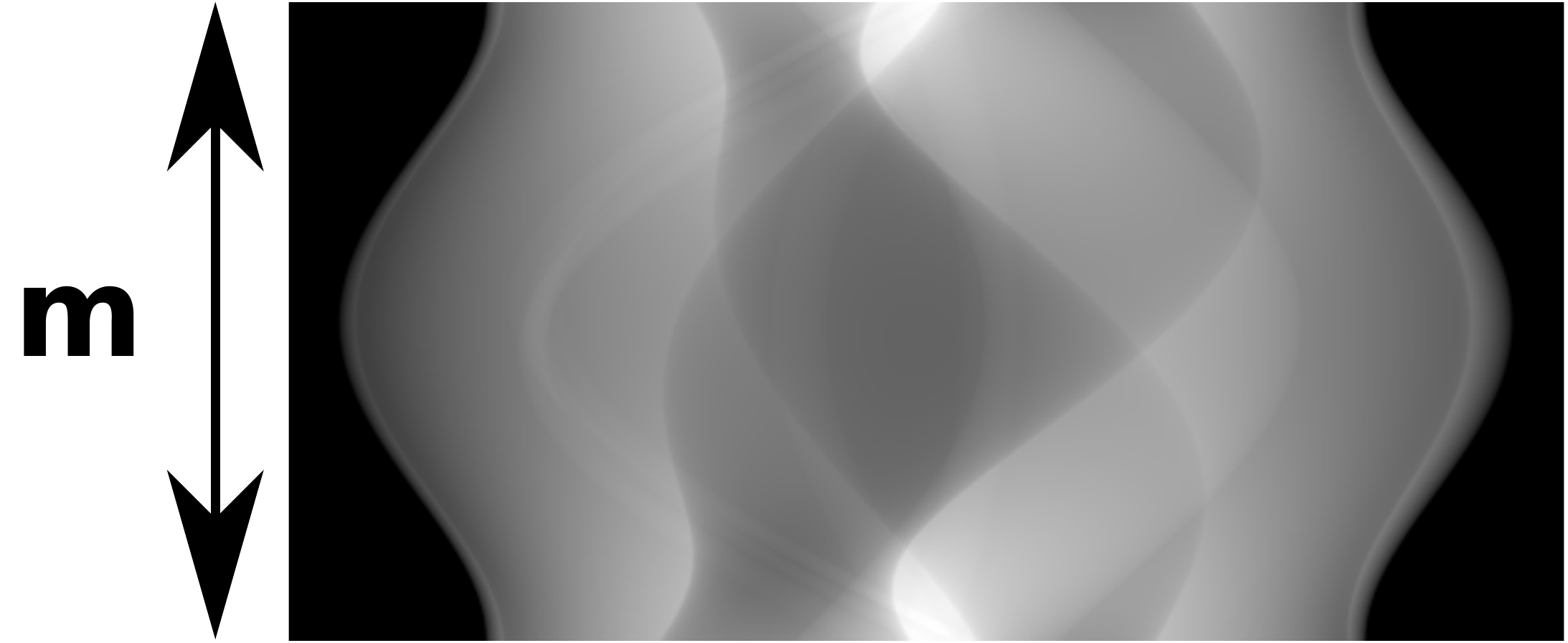}}%
  \hspace*{0.4cm}\subfloat[]{\includegraphics[width=1.2in]{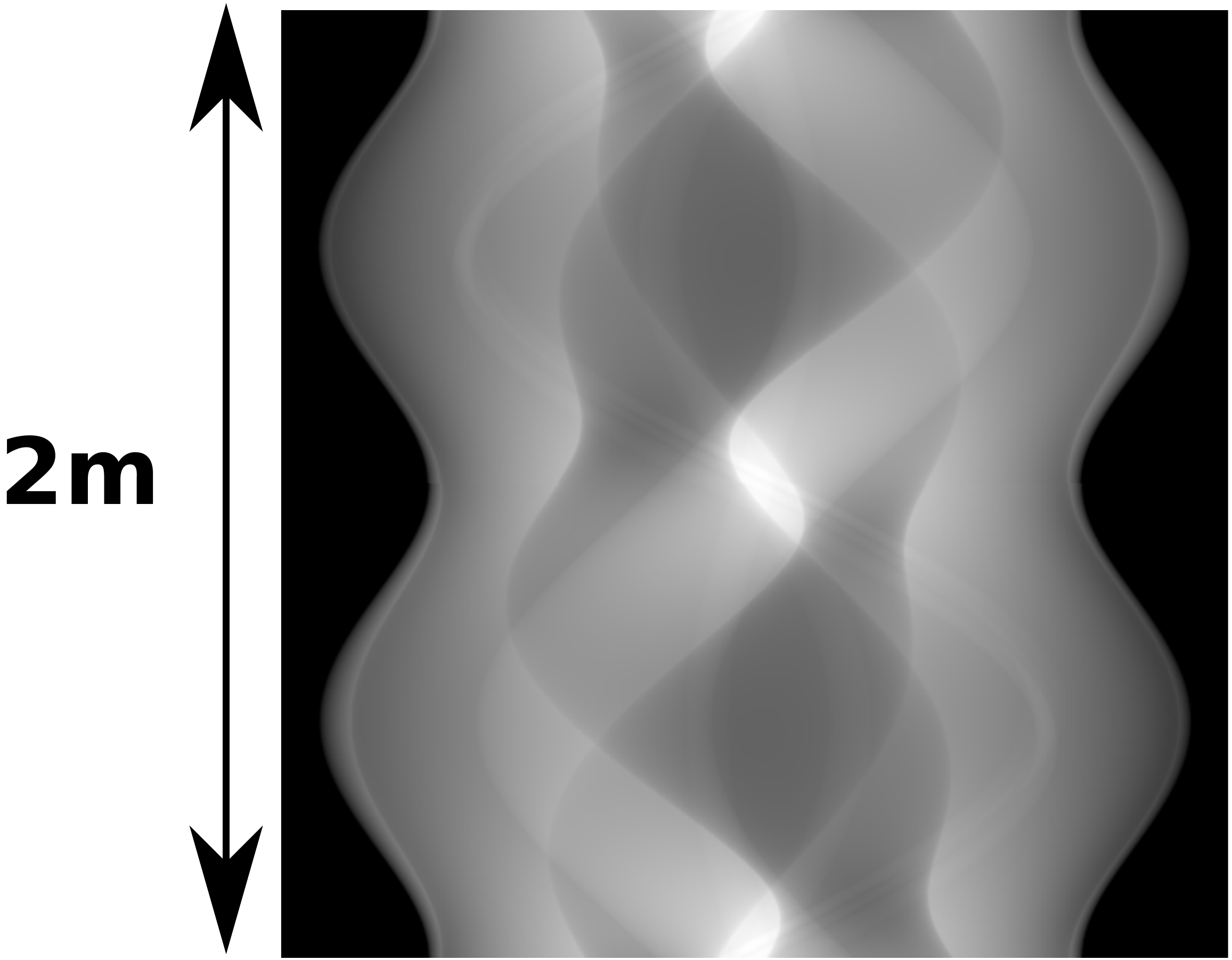}}%
  \hspace*{0.4cm}\subfloat[]{\includegraphics[width=1.2in]{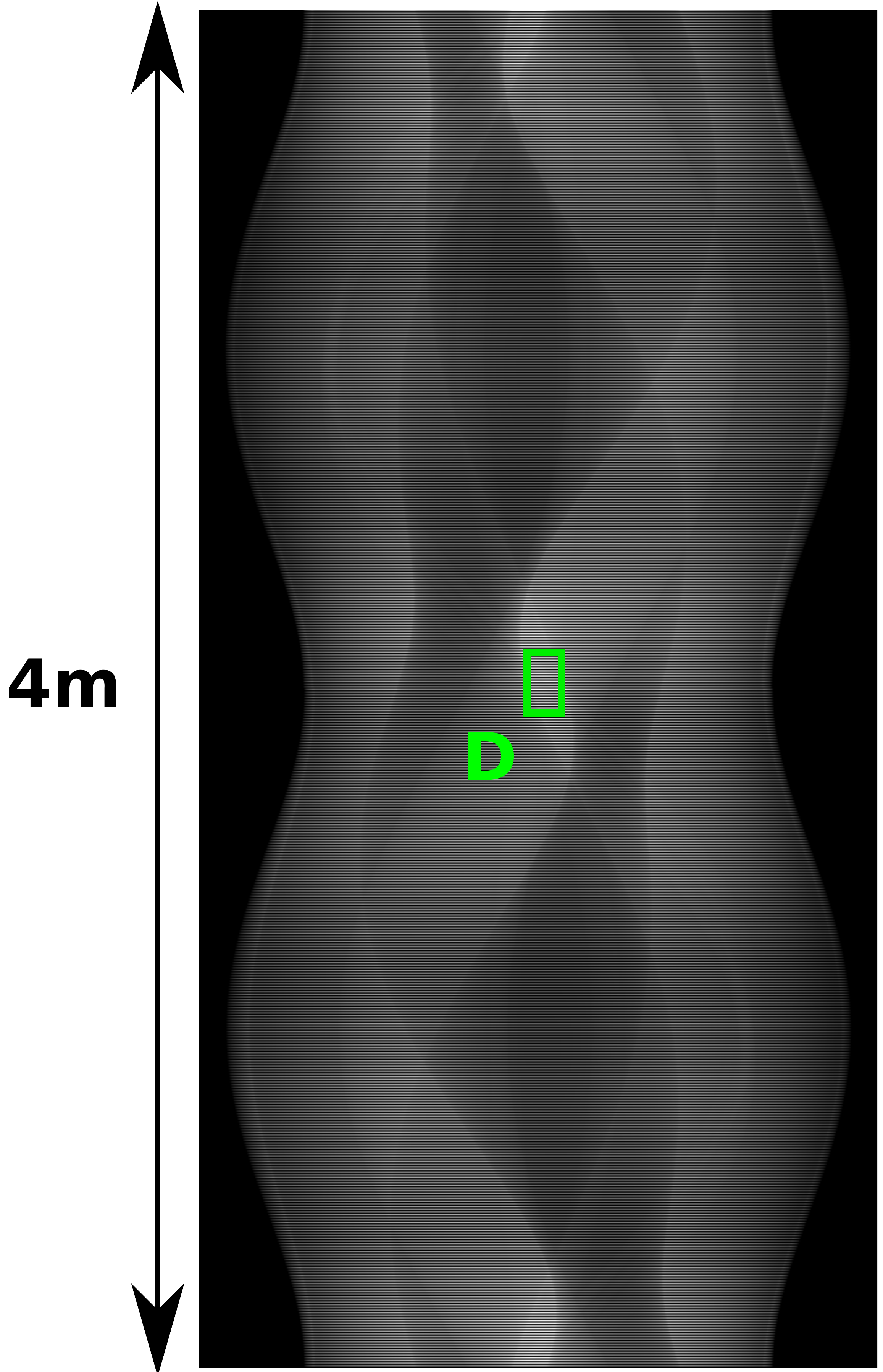}}%
  \hspace*{0.4cm}\subfloat[]{\includegraphics[width=0.8in]{fig_03_d.pdf}}
  \hspace*{0.4cm}\subfloat[]{\includegraphics[width=1.9in]{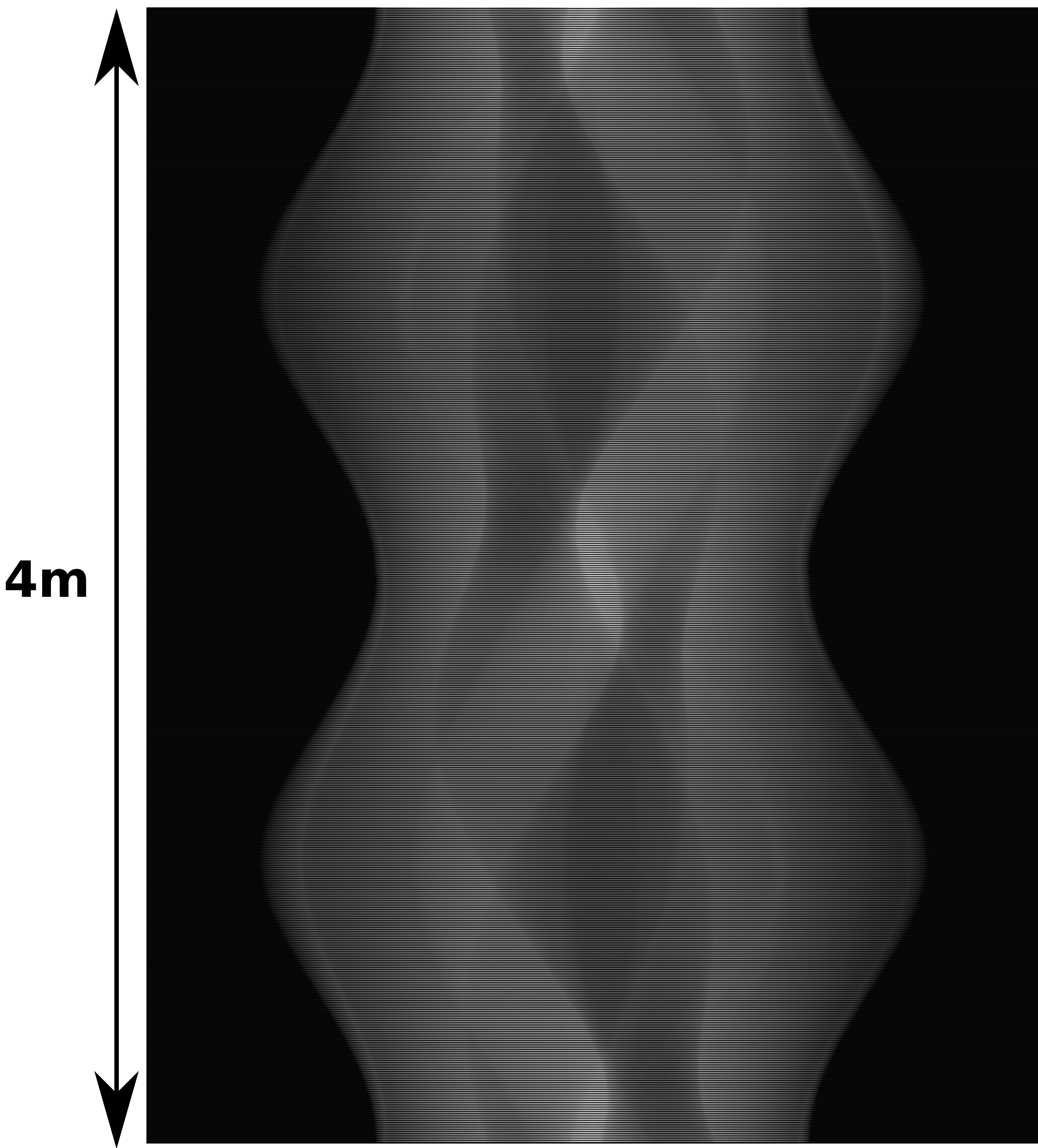}}%
  \caption{Succession of steps to compute the Fourier-Chebyshev decomposition of a sinogram.
           (a) Input sinogram of a Shepp-Logan phantom, $\theta \in [0,\pi)$. (b) Extended sinogram, $\theta \in [0,2\pi]$.
           (c) Interleaved sinogram. (d) Zoom-in of the region D in (c). (e) Sinogram with cosine-resampled
           projections.}
  \label{stages}
\end{figure}

\subsection{Step (1): extend sinogram to $[0,2\pi]$}
First, the input sinogram, $p[\theta_{h},t_{j}]$, has to be extended to $[0,2\pi]$ in order to use (\ref{fourier-chebyshev}):
\begin{equation}
  \begin{split}
	& p^{(1)}[\theta_{h'},t_{j}] = \begin{cases}
	                  p[\theta_{h},t_{j}]   & \text{for} \hspace{0.1cm} \theta_{h'} \in [0,\pi) \\[1em]
	                  p[\theta_{h},n-t_{j}] & \text{for} \hspace{0.1cm} \theta_{h'} \in [\pi,2\pi)
				      \end{cases} \hspace{0.1cm} \in \mathbb{R}^{2m \times n}\\[1em]
	& \hspace{1cm} \theta_{h'} = \frac{\pi h'}{m} \hspace{0.5cm} h' = 0,1,...,2m-1 \hspace{0.5cm}.
   \end{split}
   \label{step-1}
\end{equation}
Fig. \ref{stages}(b) shows the result of extending the example sinogram in Fig. \ref{stages}(a).

\subsection{Step (2): interleaved sinogram}
The interleaved sinogram $p^{(2)}[\theta_{h''},t_{j}]$ has $\theta_{h''} = h''\pi/2m$ for $h'' = 0,1, ...,4m-1$ and corresponds to:
\begin{equation}
  p^{(2)}[\theta_{h''},t_{j}] = \begin{cases}
				      p^{(1)}[\theta_{h'},t_{j}] & \text{for} \hspace{0.1cm} h'' = 2h' \\[1em]
				      0                   & \text{for} \hspace{0.1cm} h'' = 2h' + 1  
                                 \end{cases} \hspace{0.1cm} \in \mathbb{R}^{4m \times n}\hspace{0.3cm}.
\end{equation}
An example of interleaved sinogram is shown in Fig. \ref{stages}(c).

\subsection{Step (3): imposing HLCC}
By introducing the following quantities:
\begin{equation}
  \hspace*{0.0cm}c_{k}(\theta) = \sum\limits_{l=-\infty}^{+\infty} b_{kl}\:e^{il\theta} \hspace*{0.4cm},
\end{equation}
(\ref{fourier-chebyshev}) can be rewritten as:
\begin{equation}
  p(\theta,t) = \frac{1}{\pi} \sum\limits_{k=0}^{+\infty}c_{k}(\theta)\:(1-t^{2})^{1/2}\:U_{k}(t)\\  
  \label{fourier-chebyshev-2}
\end{equation}
The $c_{k}(\theta)$'s are the Chebyshev coefficients of $[p(\theta,t)(1-t^{2})^{-1/2}]$; 
the $b_{kl}$'s are the Fourier series coefficients of the $c_{kl}$'s.
The strategy consists in computing the $b_{kl}$ coefficients in two steps: first $p(\theta,t)$ is expanded onto the Chebyshev basis,
then, the $b_{kl}$'s are retrieved from the $c_{k}(\theta)$'s.
\newline
The $c_{k}(\theta)$'s can be easily computed,
by resampling the projections at points $t_{j}' = \cos(\pi (j+1)/(n+1))$ 
\cite{Xu2010,Bortfeld1999}. The cosine-resampling simplifies $U_{k}(t)$ into a sine function.
Thus, the first term of (\ref{fourier-chebyshev-2}) becomes:
\begin{equation}
 p^{(2)}[\theta_{h''},t_{j}'] = \sum\limits_{0}^{n-1}c_{k}[\theta_{h''}]\:\sin\left( \frac{\pi (j+1)(k+1)}{n+1}\right)\hspace*{0.3cm}.
  \label{second-step-dst}
\end{equation}
(\ref{second-step-dst}) shows that the $c_{k}[\theta_{h''}]$'s correspond to the type-1 discrete sine transform (DST-1) coefficients of
the cosine-resampled projections (an example is provided by Fig. \ref{stages}(e)).
\newline
After running the DST-1 along the channel
direction of the sinogram, the $b_{kl}$'s are finally yielded by the FFT along the view direction:
\begin{equation}
  b_{kl} = \frac{1}{4m}\:\text{DFT}\{c_{k}[\theta_{h''}]\} \hspace{0.3cm}.
  \label{second-step-dft}
\end{equation}
Altogether, imposing the HLCC on $p^{(2)}[\theta_{h''},t_{j}]$ requires the cosine resampling, a DST-1 along the channel direction,
a FFT along the view direction, setting to zero the $b_{kl}$'s according to (\ref{bkm-cond}) (Fig. \ref{checkboarded-wedge}) and reversing the process, i.e.
an IFFT along the view direction, an IDSF-1 along the channel direction and resampling at positions $t_{j}$.
The resulting sinogram is $p^{(3)}[\theta_{h''},t_{j}] \in \mathbb{R}^{4m \times n}$.

\subsection{Step (4): crop in $[0,\pi)$ and reassign original data}
The sinogram is, finally, cropped again in the interval $[0,\pi)$ and the original projections are reassigned.


\section{Complexity and efficiency}
\label{complexity}
The computational cost of the HLSF lies in the calls of the DST-1 along the channel direction and the FFT along
the view direction. The computation of the DST can be factorized similarly to the FFT 
($O(N\log_{2}N)$ complexity) plus few pre- and post-processing steps with $O(N)$
complexity \cite{Poularikas2000}. For an input sinogram with $m$ views $\times$ $n$ pixels, this yields to
approximately $8mn(\log_{2}n + \log_{2}4m)$ floating operations and a resulting $O(mn(\log_{2}n + \log_{2}4m))$
complexity.
\newline
To show that the proposed filter only slightly impacts the total reconstruction speed, 
a pure Matlab implementation of the HLSF has been compared 
to the well known Matlab function \textit{iradon}.
Results for sinograms of different sizes, collected in Tab. \ref{tab-comparison-times},
prove that the HLSF requires smaller runtimes than a standard non-GPU implementation of FBP,
especially for real datasets where $m,n > 10^{3}$.
\begin{table}[!h]\centering
    \centering
    \ra{1.3}\hspace*{0.0cm}
    \begin{tabular}{@{}rrrrcrrrcrrr@{}}\toprule
	    & HLSF & FBP  \\ \midrule
	805 views $\times$ 512 pix. & 0.22 s & 0.30 s & \\ \midrule
	1608 views $\times$ 1024 pix. & 0.63 s & 2.34 s &  \\ \midrule
        2500 views $\times$ 2048 pix. & 1.65 s & 14.50 s &  \\ 	
	\bottomrule
    \end{tabular}
  \caption{Comparison between the time elapsed to run the HLSF and the FBP reconstruction for sinograms
            of different sizes. HLSF is implemented in pure Matlab.
            FBP is performed by the Matlab function \textit{iradon}.}
  \label{tab-comparison-times}
\end{table}


\section{Benchmark procedure}
\label{benchmark-procedure}
\begin{figure}[!t]
  \centering
  \hspace*{-0.8cm}\subfloat[PH-1]{\includegraphics[width=1.5in]{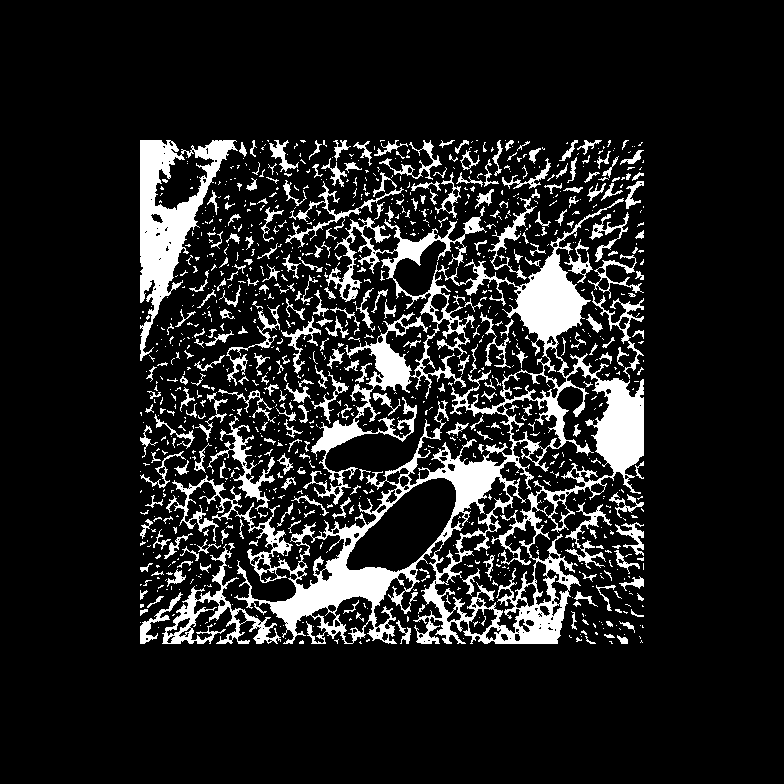}\label{}}
  \hspace*{0.4cm}\subfloat[PH-2]{\includegraphics[width=1.5in]{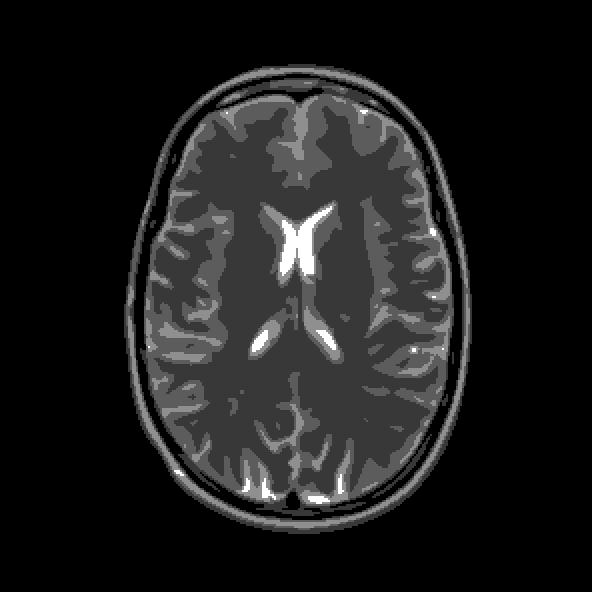}\label{}}
  \hspace*{0.4cm}\subfloat[PH-3]{\includegraphics[width=1.5in]{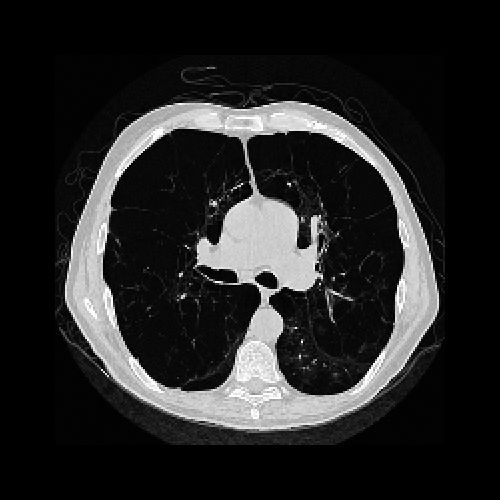}\label{}}
  \hspace*{0.4cm}\subfloat[PH-4]{\includegraphics[width=1.5in]{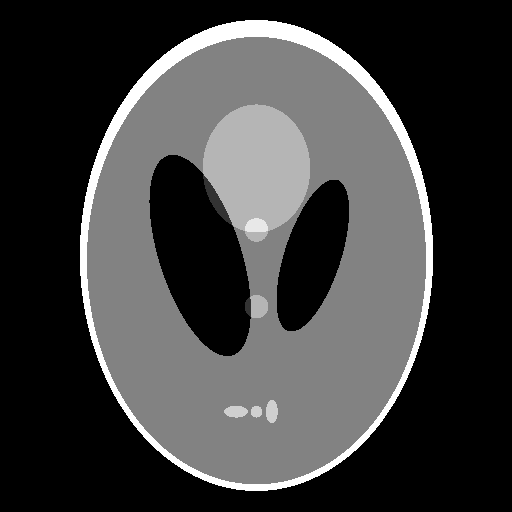}\label{}}
  \caption{Set of simulated data used to benchmark the HLSF. PH-1 has 784 $\times$ 784 pixels;
           PH-2 has 592 $\times$ 592 pixels; PH-3 has 500 $\times$ 500 pixels; PH-4 has 512 $\times$ 512 pixels.}
  \label{phantom-set}
\end{figure}

To assess the performance of the HLSF, four phantoms
with structural patterns of different complexity have been considered (Fig. \ref{phantom-set}). 
PH-1 in Fig. \ref{phantom-set}(a) is the segmentation of 
a reconstructed slice of mouse lung tissue at micrometer scale.
PH-2 in Fig. \ref{phantom-set}(b) is a multilevel segmentation of a MRI scan of a human brain.
PH-3 in Fig. \ref{phantom-set}(c) is a multilevel segmentation of a CT slice of a human lung.
PH-4 is the well-known Shepp-Logan phantom \cite{Shepp1974}.
\newline
The simulated sinograms are computed by a standard space-based implementation of
the Radon transform based on slant-stacking with linear interpolation \cite{Toft1996}.
\newline
Reconstructions are performed by means of FBP. The tradeoff between signal-to-noise ratio (SNR)
and spatial resolution of FBP reconstructions is highly dependent on the choice of the filter function. For this reason, the dataset
has been reconstructed with 3 different filters: a pure ramp or Ram-Lak filter and a ramp combined to a Hanning
or Parzen \cite{Lyra2011} window to damp the high frequency components of the projections \cite{Kak2001}.
These filters are indicated in the 
following as Ram-Lak, Hanning and Parzen, respectively.
The Ram-Lak filter provides the highest spatial resolution and the worst SNR, the opposite occurs 
for Parzen; Hanning is placed in the middle. Reconstructions labelled ``CFBP'' (Consistent FBP)
were computed with FBP on a sinogram pre-filtered by HLSF with an upsampling factor of 2;
reconstructions labelled ``IFBP'' (Interpolation FBP) were computed with FBP on a sinogram upsampled by a factor of 
2 by means of 1D cubic spline interpolation along the view direction;
otherwise they are simply labelled ``FBP''.
\newline
The standard peak-signal-to-noise ratio (PSNR) \cite{HuynhThu2008}, calculated within the reconstruction circle,
is used to score each reconstruction with respect to the corresponding phantom. 
\newline
The ``sampling factor'' (SF) is defined as the ratio between the number of projections of the considered sinogram
and the number of projections of an optimally sampled sinogram. For parallel beam geometry, a sinogram is optimally sampled if $m \geq n \:\pi/2$, 
with $m$ the number of views and $n$ the number of detector pixels \cite{Kak2001}. A sinogram with 100 views $\times$
512 pixels, for example, considered well-sampled with $805 \backsimeq 512 \cdot \pi/2$ views, has SF = 0.12.

\section{Experiments}
\label{experiments}
\begin{figure}[!b]
\centering
    \hspace*{0.0cm}\subfloat[Ram-Lak]{\includegraphics[width=3.0in]{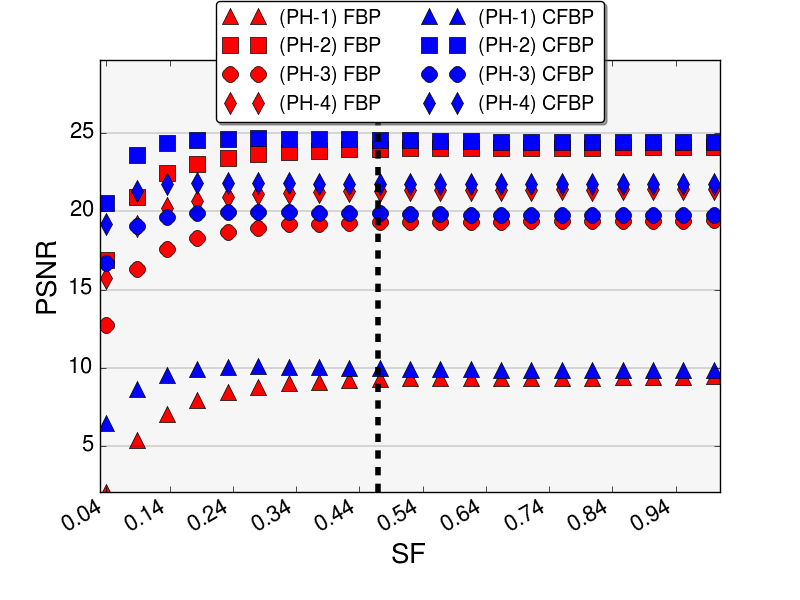}\label{}}%
    \hspace*{1.0cm}\subfloat[Hanning]{\includegraphics[width=3.0in]{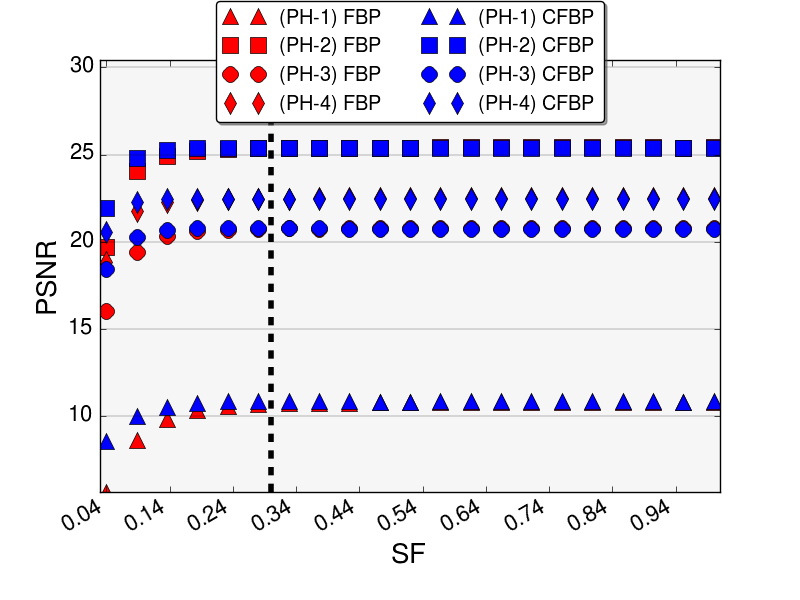}\label{}}\\
    \hspace*{0.0cm}\subfloat[Parzen]{\includegraphics[width=3.0in]{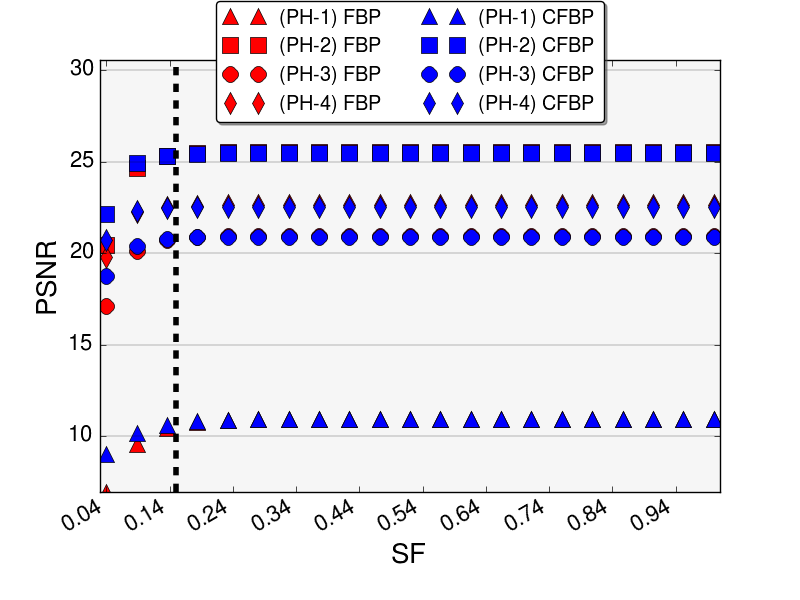}\label{}}
  \caption{PSNR scores as a function of the sampling factor, SF, of the 
           FBP reconstructions performed with Ram-Lak, Hanning and Parzen filters. The red markers
           correspond to FBP reconstructions, the blue ones to CFBP reconstructions.
           The marker shape is related to the phantom used to create the simulated data.}
  \label{exp-down-psnr}
\end{figure}

\begin{figure}[!t]
\centering
    \hspace*{0.0cm}\subfloat[Ram-Lak (PH-1)]{\includegraphics[width=3.0in]{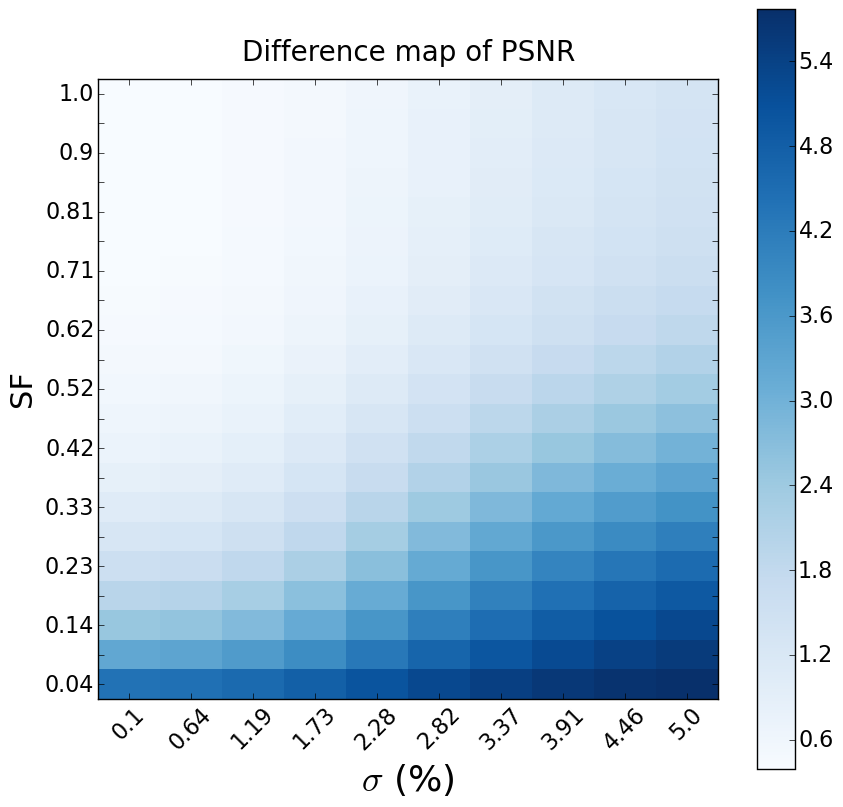}\label{}}%
    \hspace*{1.0cm}\subfloat[Hanning (PH-2)]{\includegraphics[width=3.0in]{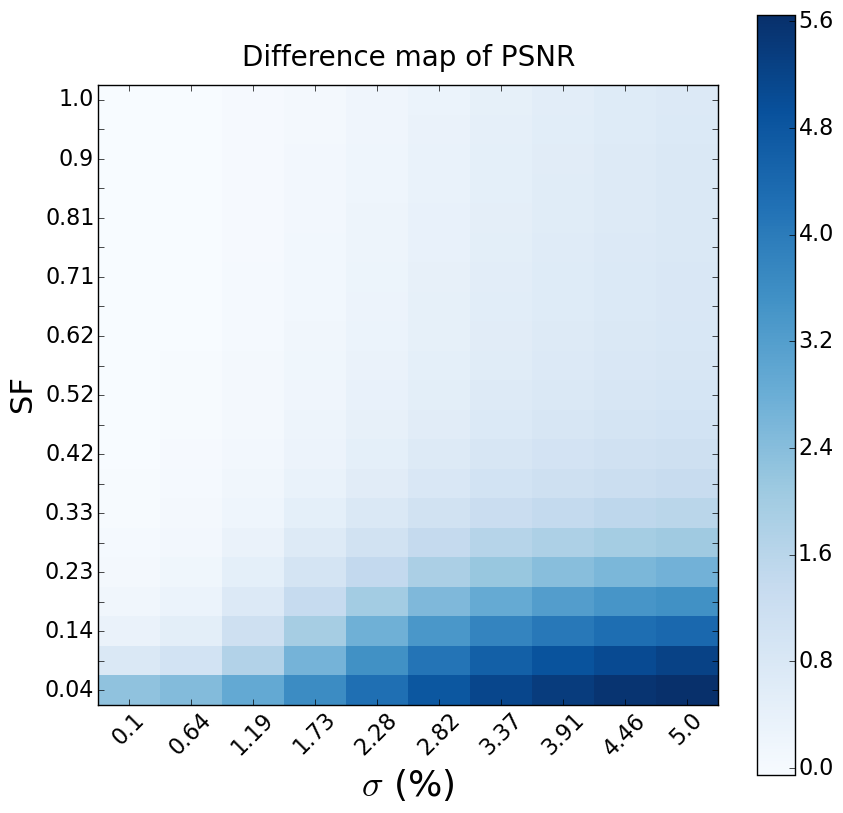}\label{}}\\
    \hspace*{0.0cm}\subfloat[Parzen (PH-3)]{\includegraphics[width=3.0in]{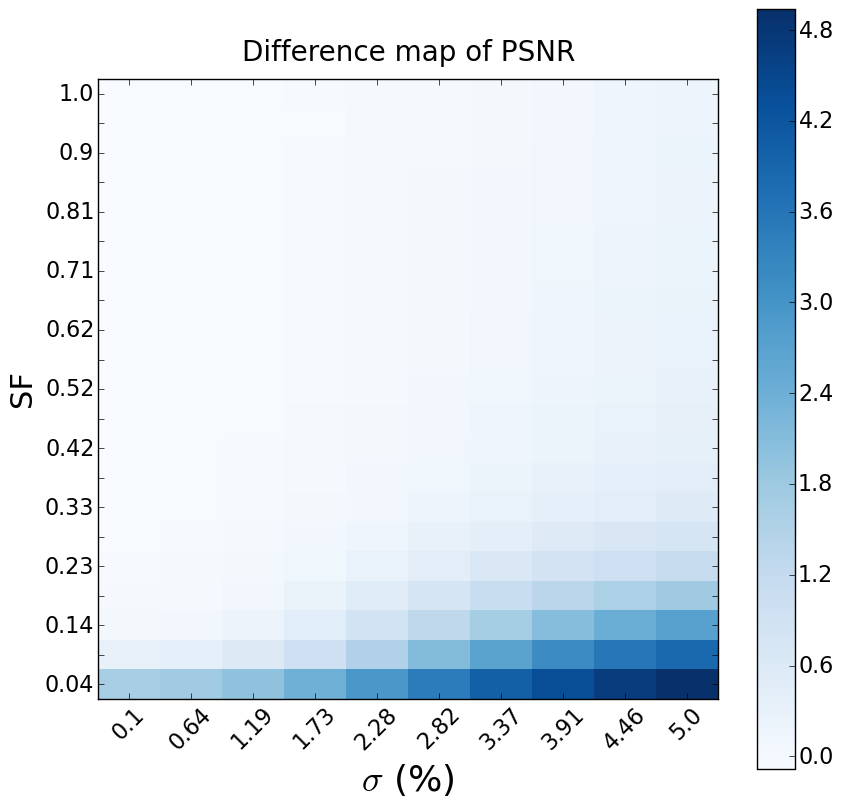}\label{}}%
  \caption{Maps of difference between PSNR scores of the CFBP and the FBP reconstructions as a function of the sampling factor, SF, and the variance, $\sigma$,
           of the additional Poisson noise. $\sigma$ is expressed as percentage of the original sinogram mean value.
           The caption of each map specifies what filter and phantom were used 
           for the reconstructions.}
  \label{exp-down-noise-psnr}
\end{figure}

\begin{figure}[!t]
\centering
  \hspace*{0cm}\subfloat[FBP]{\includegraphics[width=3.0in]{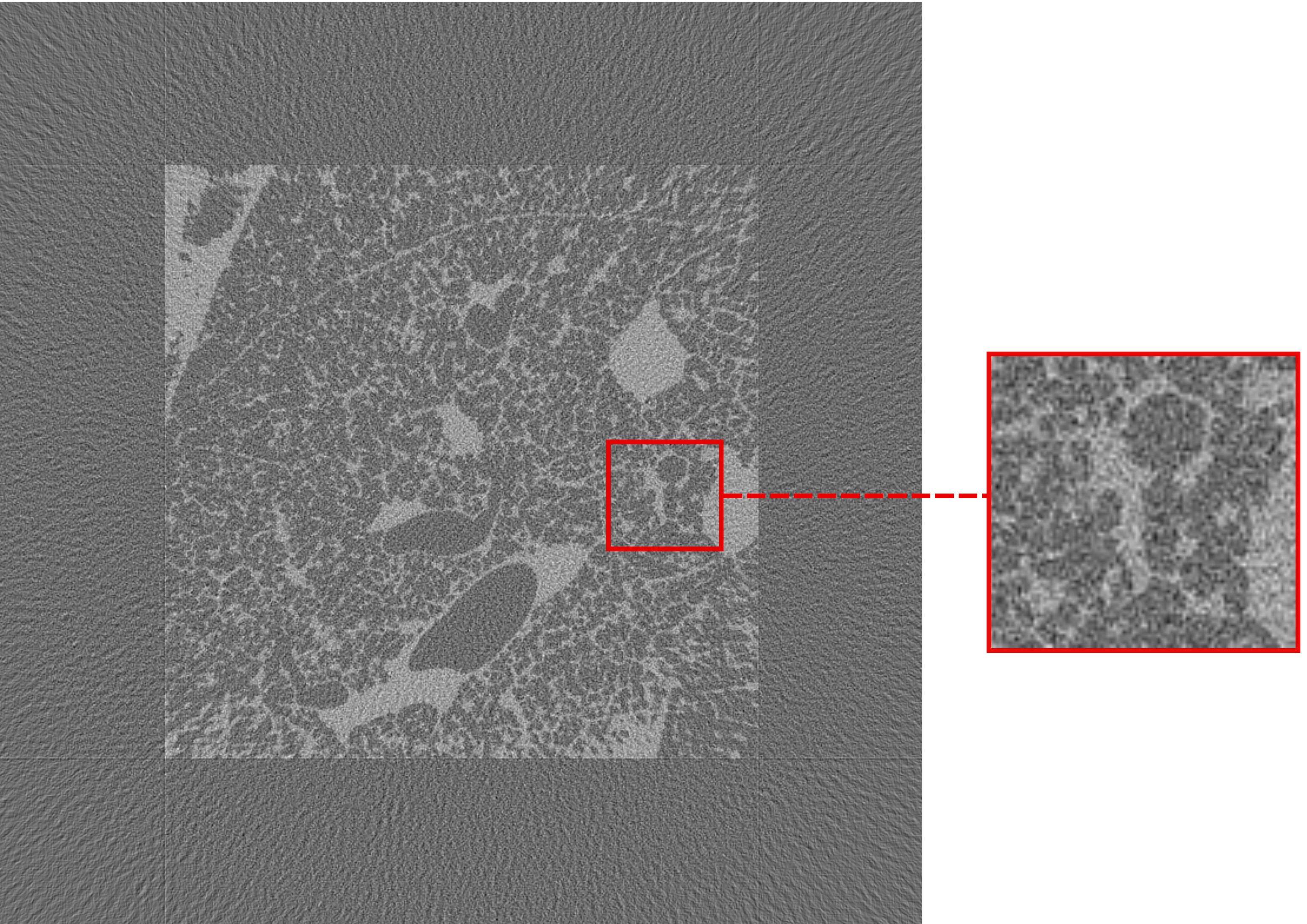}\label{}}%
  \hspace*{0.5cm}\subfloat[CFBP]{\includegraphics[width=3.0in]{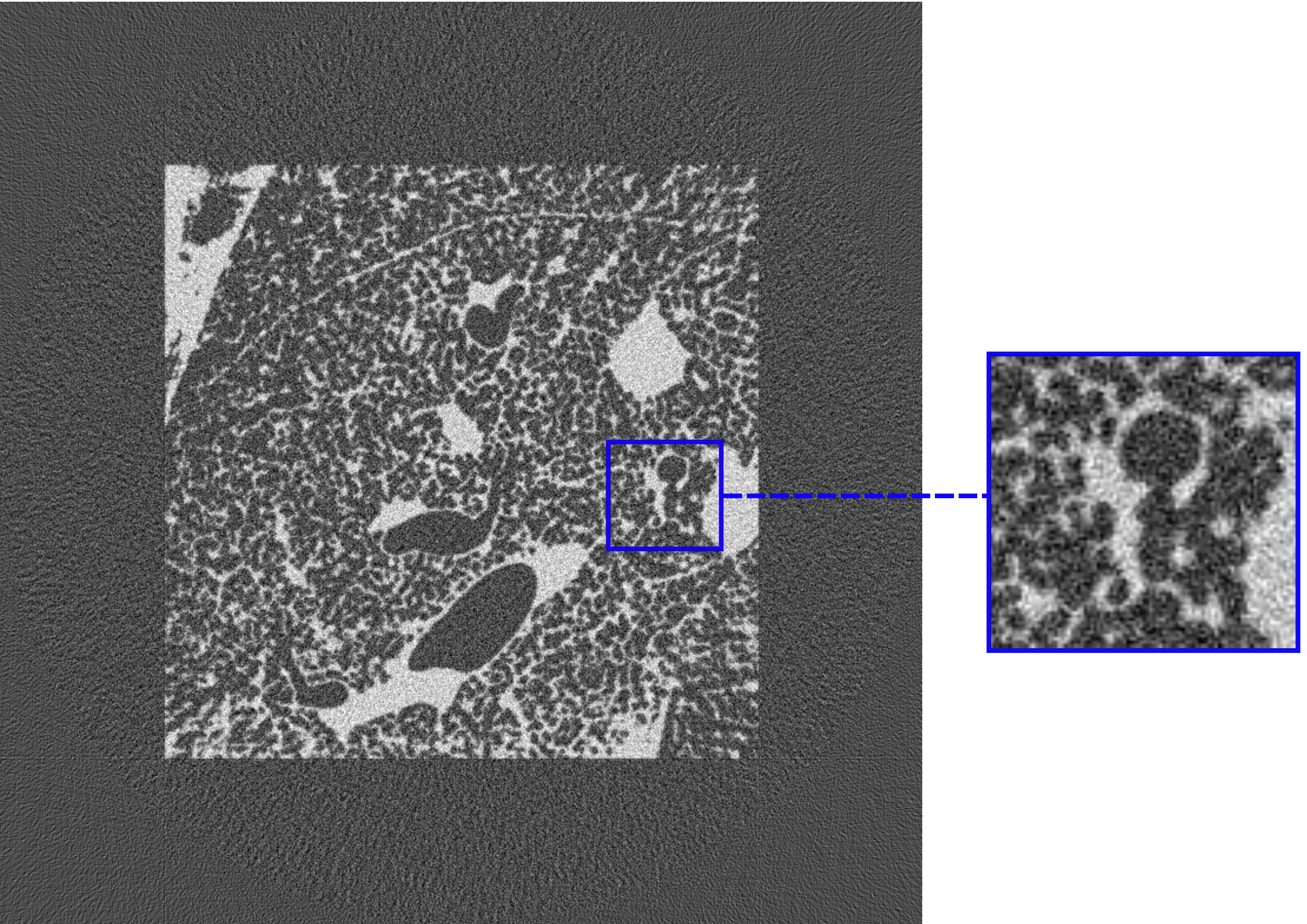}\label{}}
  \caption{Reconstructions performed by FBP and CFBP with Ram-Lak filter of a PH-1 sinogram 168 views $\times$ 784
           pixels (SF = 14\%) + Poisson noise with $\sigma = 2.2\%$ of the original sinogram mean value.}
  \label{reco-1}
\end{figure}

\begin{figure}[!t]
\centering
  \hspace*{0cm}\subfloat[FBP]{\includegraphics[width=3.0in]{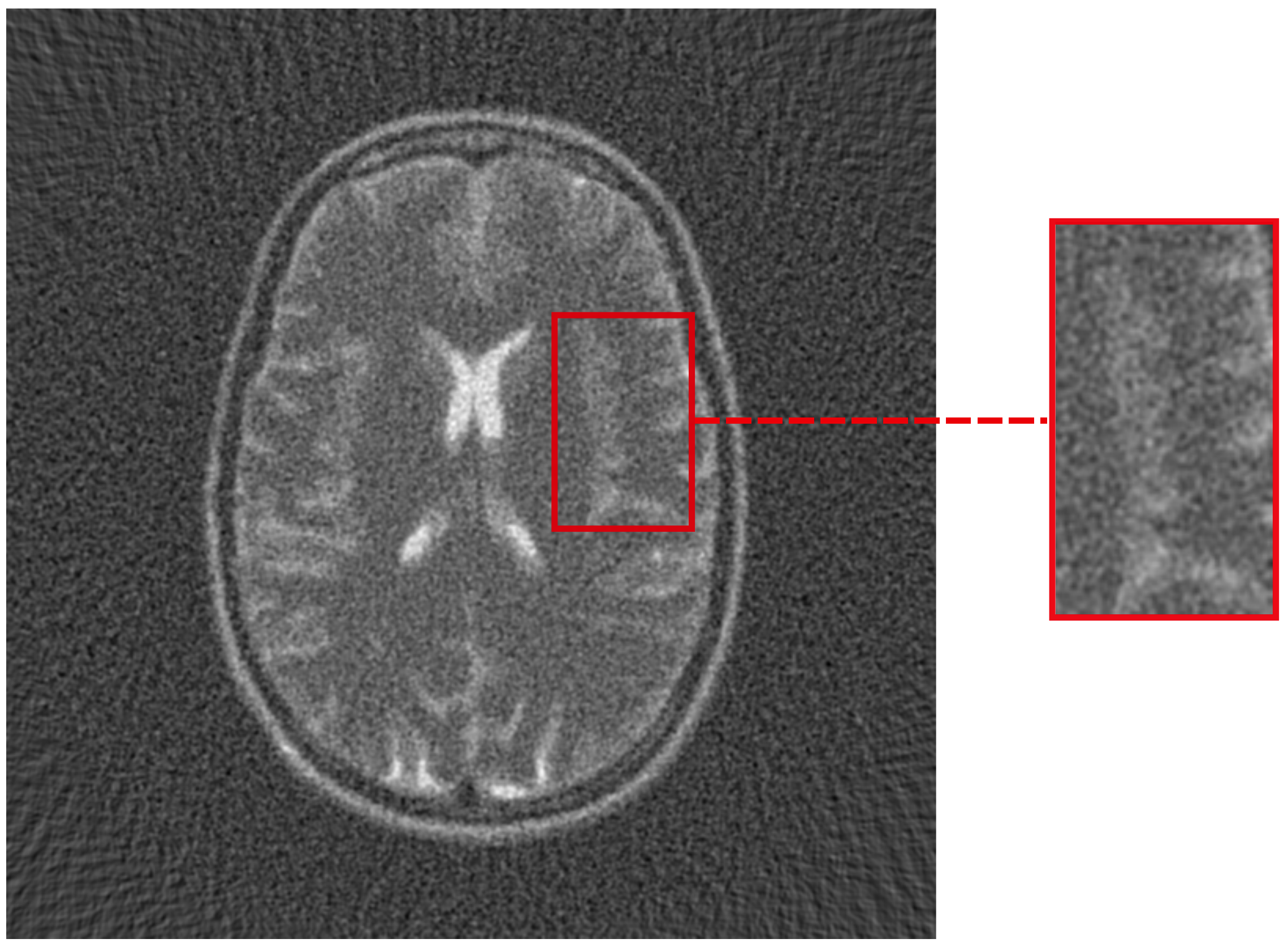}\label{}}%
  \hspace*{0.5cm}\subfloat[CFBP]{\includegraphics[width=3.0in]{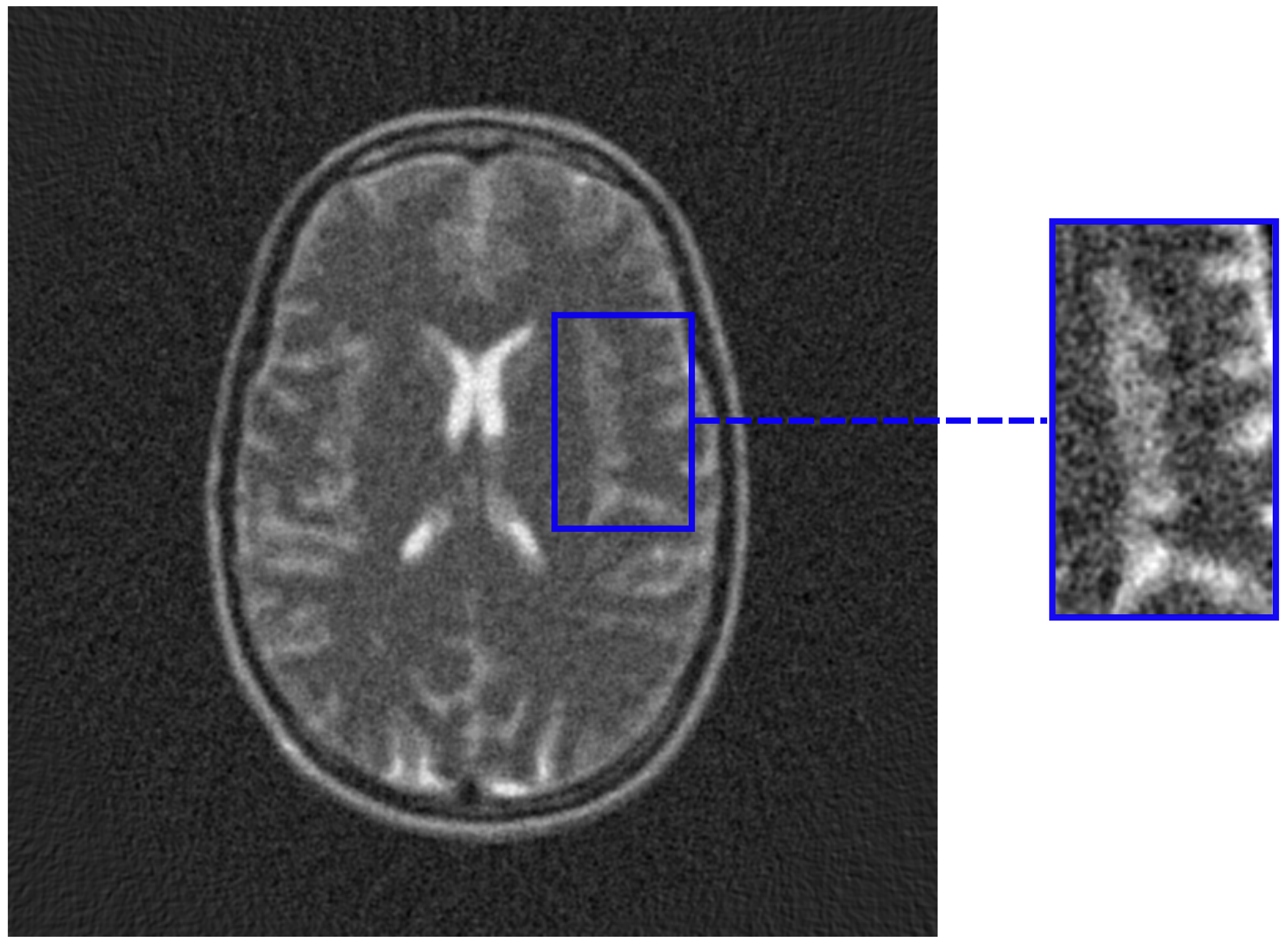}\label{}}
  \caption{Reconstructions performed by FBP and CFBP with Parzen filter of a PH-2 sinogram 82 views $\times$ 592 pixels (SF = 9\%)
           + Poisson noise with $\sigma = 2.8\%$ of the original sinogram mean value.}
  \label{reco-2}
\end{figure}

\begin{figure}[!t]
\centering
    \hspace*{0cm}\subfloat[FBP]{\includegraphics[width=3.0in]{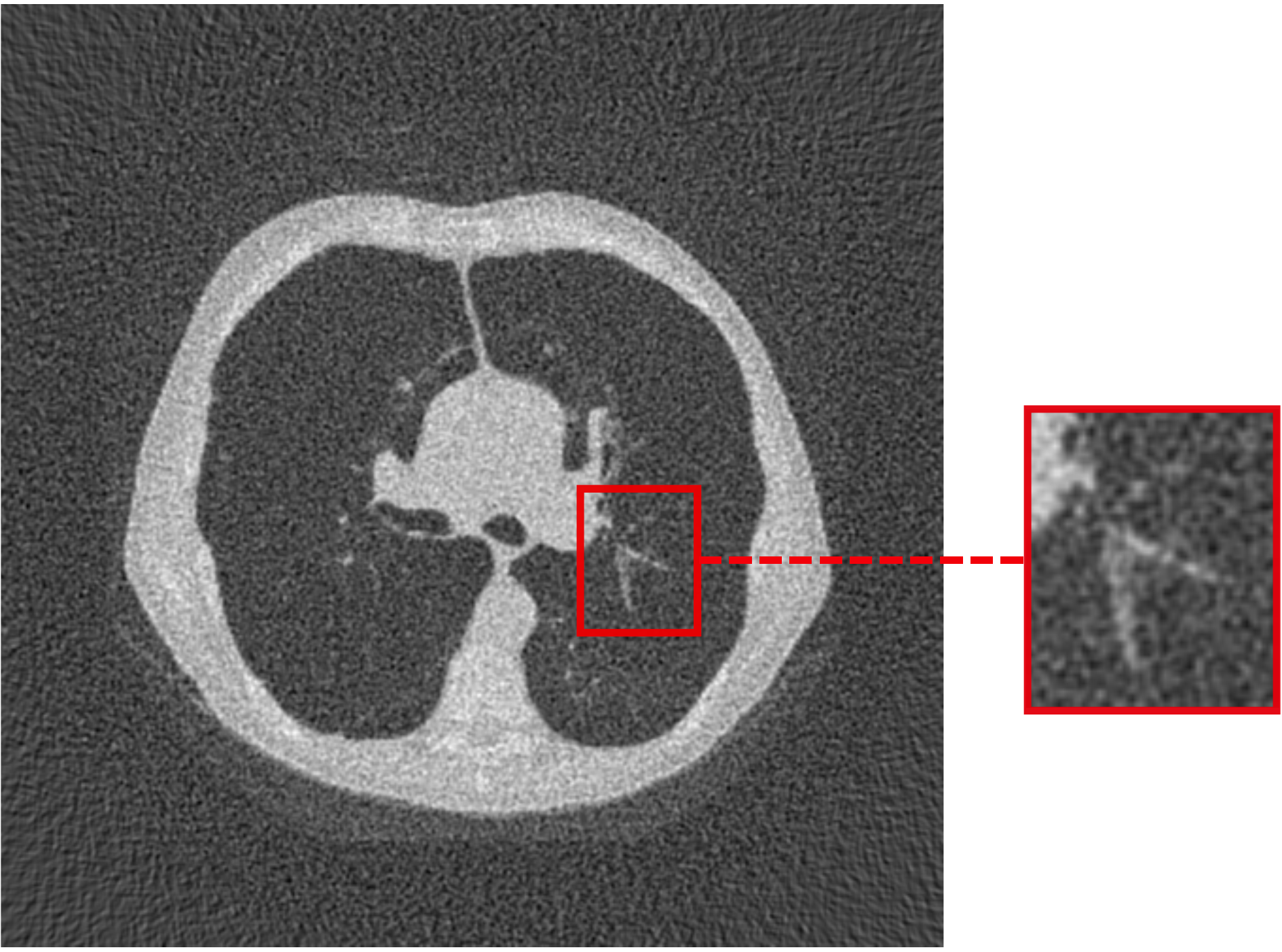}\label{}}%
    \hspace*{0.5cm}\subfloat[CFBP]{\includegraphics[width=3.0in]{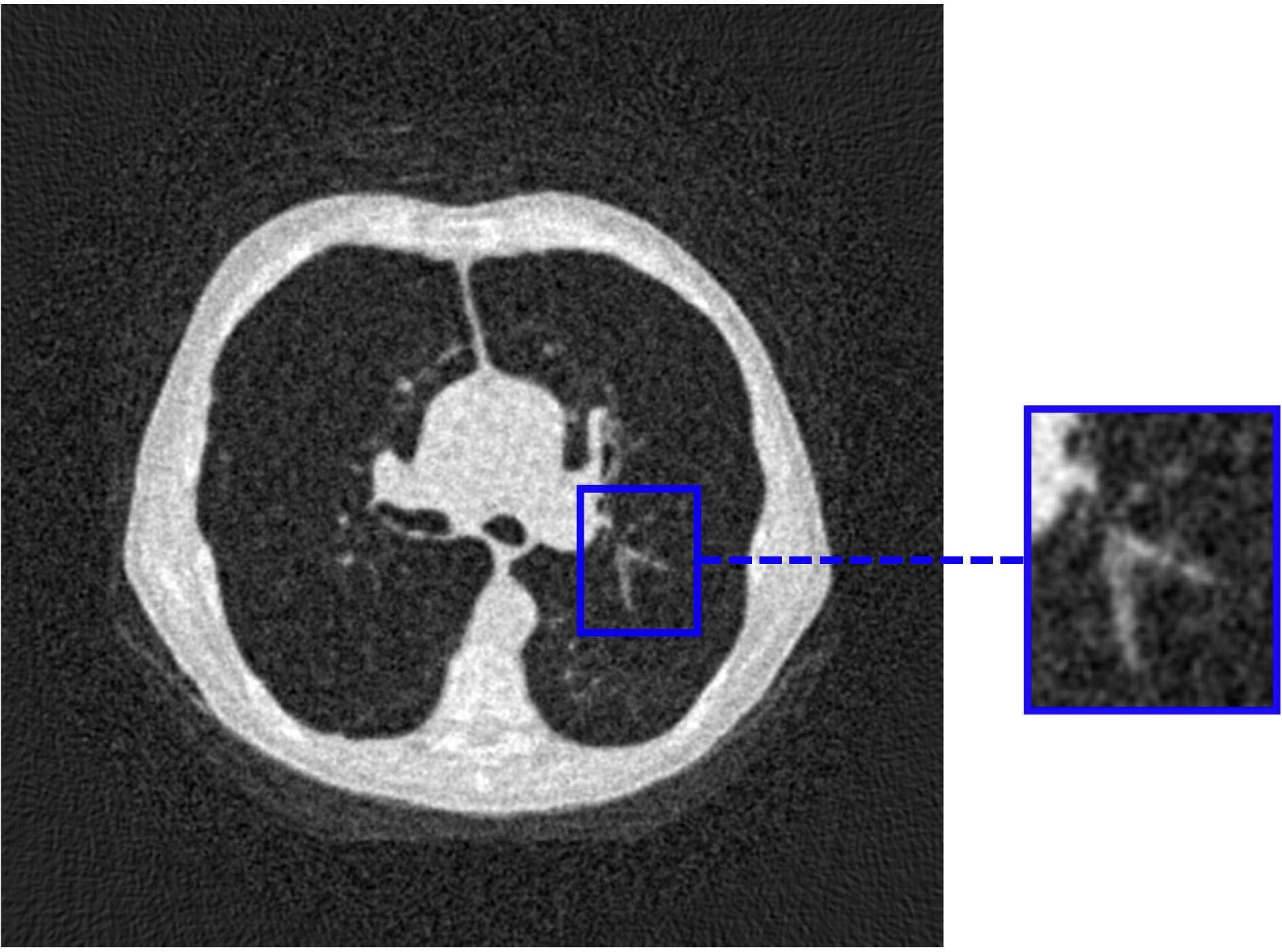}\label{}}\\ 
  \caption{Reconstructions performed by FBP and CFBP with Hanning filter of a PH-3 sinogram 108 views $\times$ 500 pixels (SF = 14\%)
           + Poisson noise with $\sigma = 2.2\%$ of the original sinogram mean value. }
  \label{reco-3}
\end{figure}

\begin{figure}[!t]
\centering
    \hspace*{0cm}\subfloat[FBP]{\includegraphics[width=3.0in]{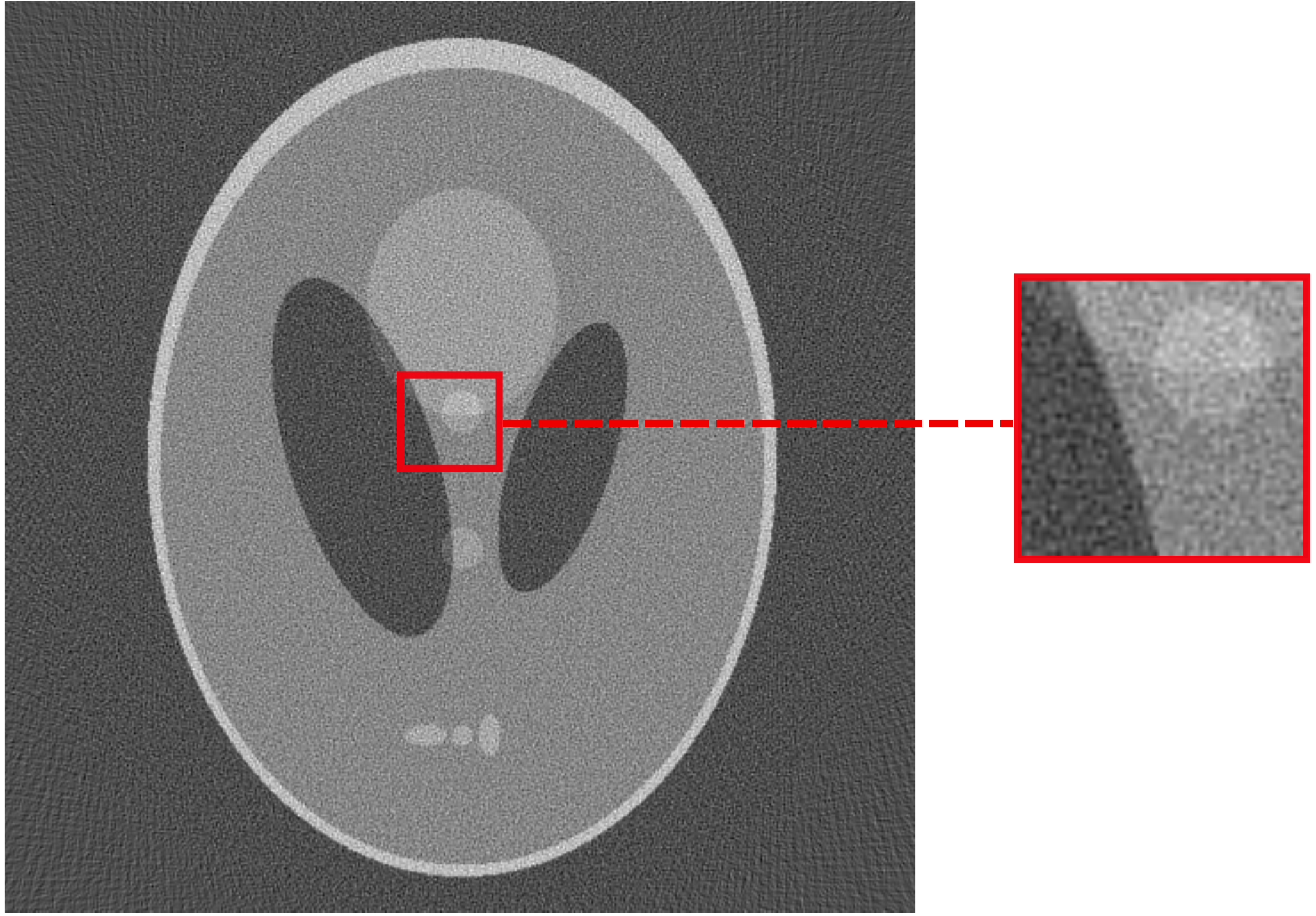}\label{}}%
    \hspace*{0.5cm}\subfloat[CFBP]{\includegraphics[width=3.0in]{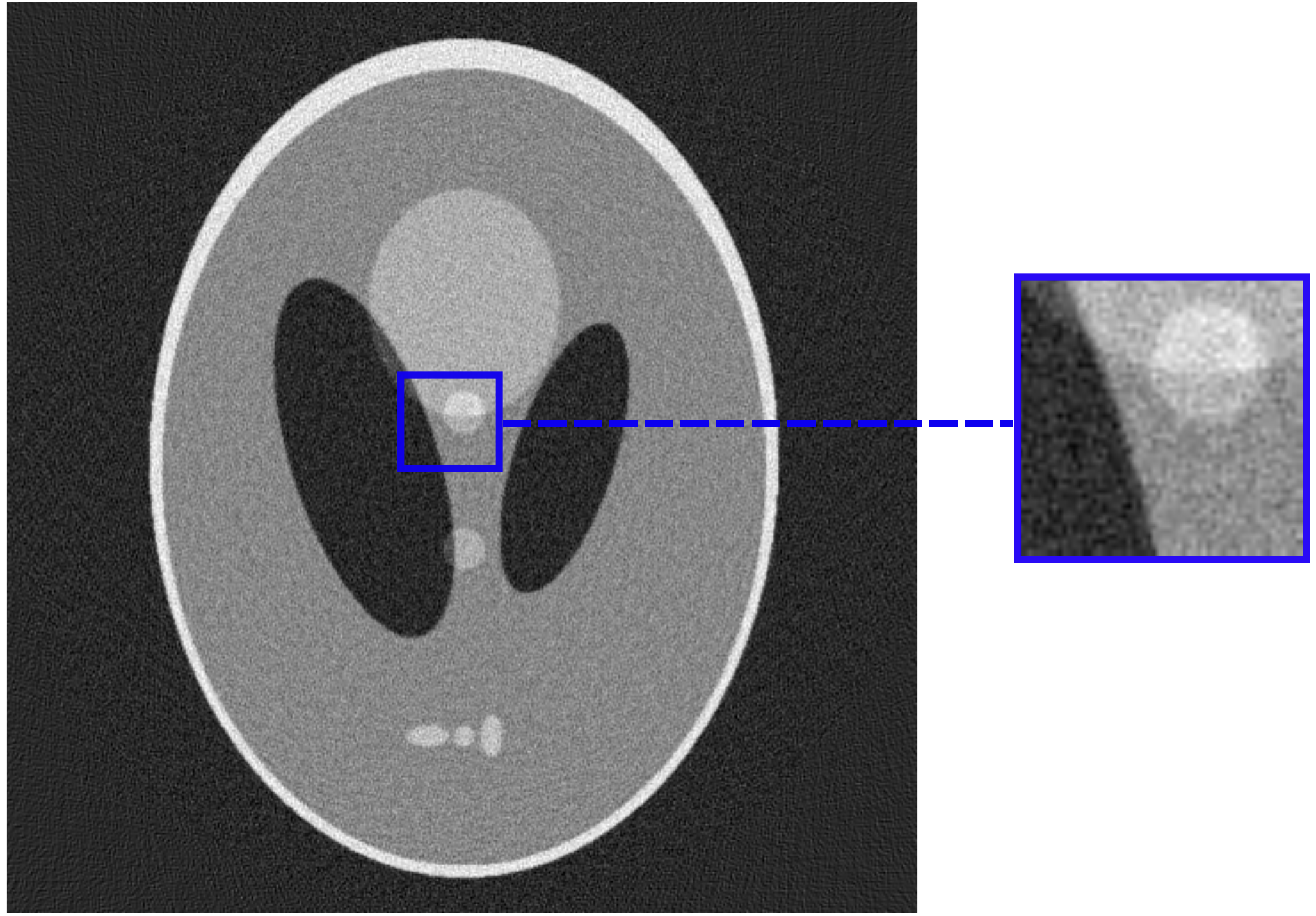}\label{}}
  \caption{Reconstructions performed by FBP and CFBP with Ram-Lak filter of
           a PH-4 sinogram 149 views $\times$ 512 pixels (SF = 18\%) + Poisson noise with $\sigma = 1.1\%$ of the original sinogram mean value.}
  \label{reco-4}
\end{figure}

First, the performance of HLSF is tested for the noiseless case: noise-free sinograms with different SFs are
reconstructed with FBP after pre-filtering with HLSF.
The plots in Fig. \ref{exp-down-psnr} show that HLSF improves the reconstruction quality for small SF,
while for higher SF the results for FBP and CFBP are comparable. 
The exact boundary between these 2 regimes depends on the filter choice.
It is marked in Fig. \ref{exp-down-psnr} with a dashed black line and corresponds roughly to 0.47 for the Ram-Lak filter, 
0.30 for Hanning and 0.15 for Parzen, regardless of the reconstructed object. 
The stronger the action of the filter, the smaller the SF interval where the HLSF increases the 
PSNR of the reconstruction. 
\newline
Subsequently, HLSF is also tested for noisy sinograms.
The standard deviation of the added Poisson noise is expressed as percentage of the original sinogram mean value
and is indicated with $\sigma$.
Fig. \ref{exp-down-noise-psnr} presents two-dimensional maps showing the difference between the PSNR score of CFBP and FBP reconstructions.
Positive values indicate that CFBP outperforms FBP and viceversa.
Each map corresponds to a specific choice of reconstruction filter and phantom. 
\newline
The differential maps, shown in Fig. \ref{exp-down-noise-psnr}, are characterized by the same trend observed for the PSNR
in Fig. \ref{exp-down-psnr}: the stronger the action of the filter, the smaller the area, where HLSF provides 
a substantial improvement. However, CFBP reconstructions have always a higher PSNR compared to the FBP ones,
since values on these maps are everywhere positive (Fig. \ref{exp-down-noise-psnr}). 
\newline
Reconstructions with FBP and CFBP for each phantom are displayed in Fig. \ref{reco-1}, \ref{reco-2}, \ref{reco-3} and \ref{reco-4}.


\begin{figure}[!t]
\centering
    \hspace*{0.0cm}\subfloat[Ram-Lak]{\includegraphics[width=3.0in]{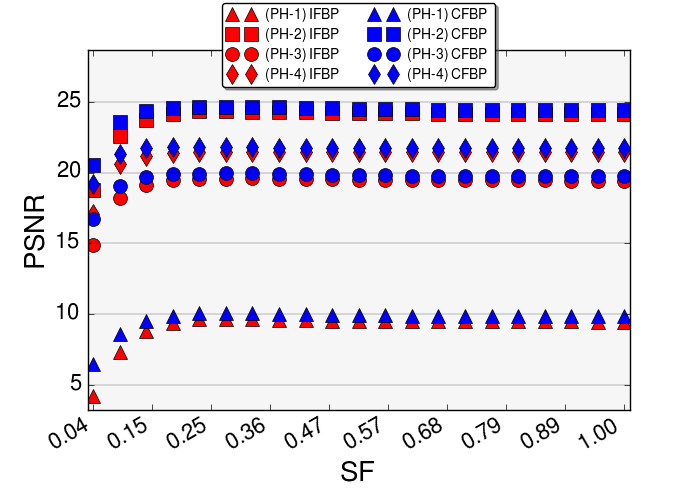}\label{}}%
    \hspace*{1.0cm}\subfloat[Hanning]{\includegraphics[width=3.0in]{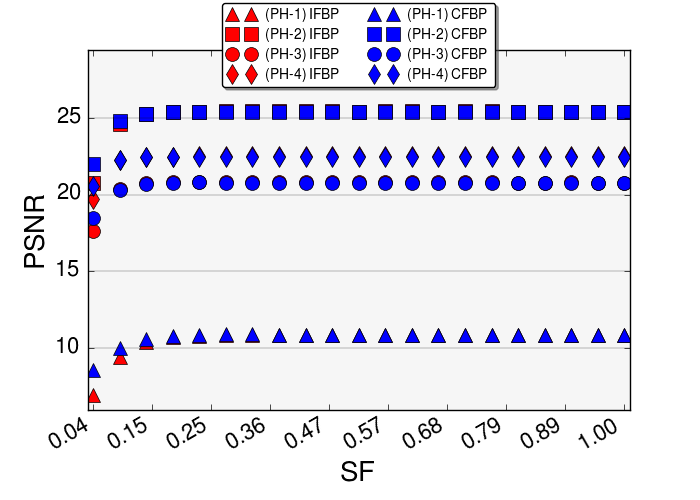}\label{}}\\
  \caption{PSNR scores as a function of the sampling factor, SF, of the 
           FBP reconstructions performed with Ram-Lak and Hanning filters. The red markers
           correspond to IFBP (cubic spline interpolation along the view direction + FBP) reconstructions, 
           the blue ones to CFBP reconstructions.
           The marker shape is related to the phantom used to create the simulated data.}
  \label{exp-down-psnr-cubic}
\end{figure}
\begin{figure}[!t]
\centering
    \hspace*{-1.0cm}\subfloat[Ram-Lak (PH-4)]{\includegraphics[width=3.0in]{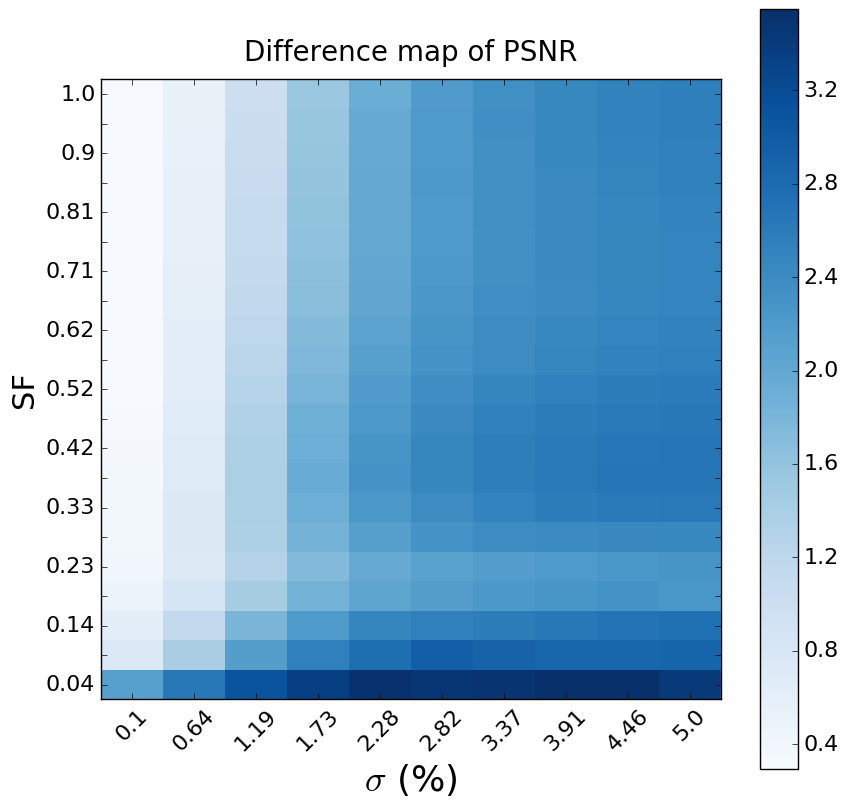}\label{}}%
    \hspace*{1.0cm}\subfloat[Hanning (PH-4)]{\includegraphics[width=3.0in]{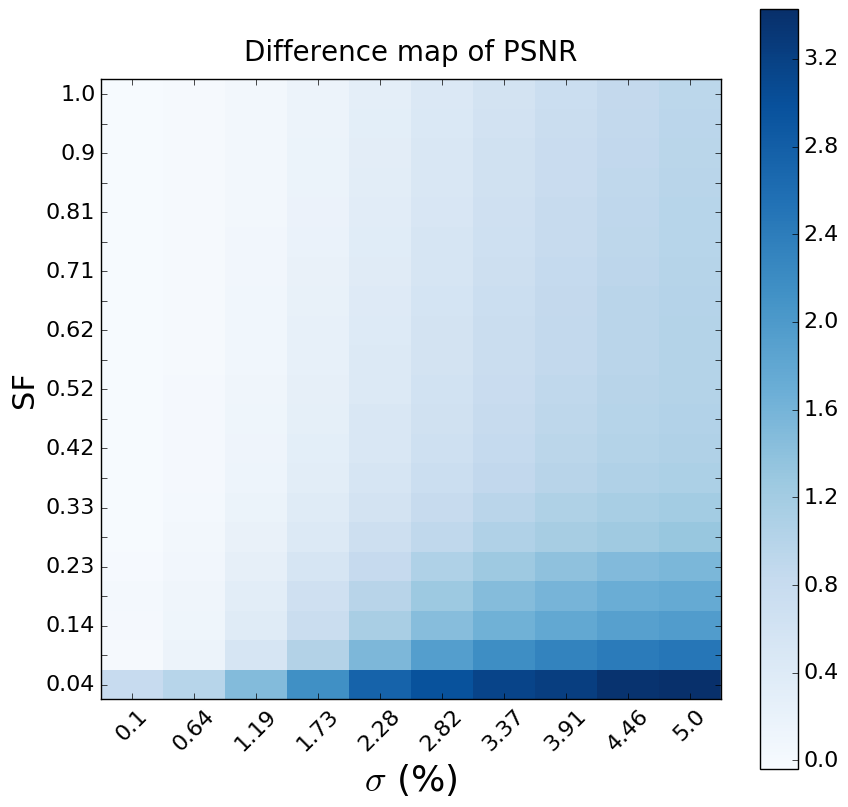}\label{}}\\
    \hspace*{-0.5cm}\subfloat[Parzen (PH-4)]{\includegraphics[width=3.0in]{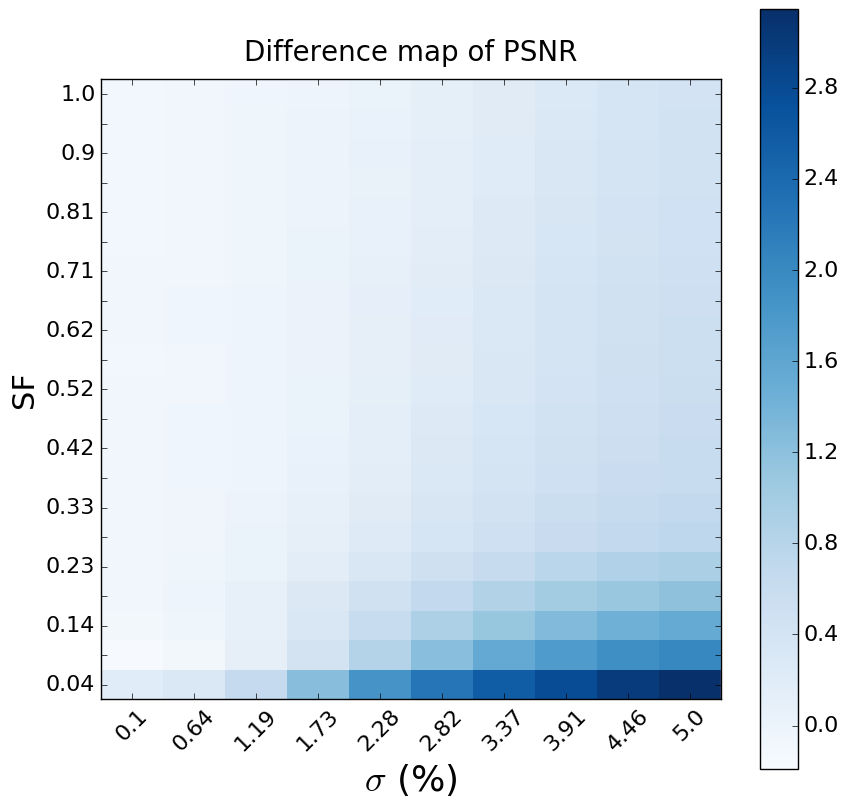}\label{}}%
  \caption{Maps of difference between PSNR scores of the CFBP and the IFBP (cubic spline interpolation along the view direction + FBP)
           reconstructions as a function of the sampling factor, SF, and the variance, $\sigma$,
           of the additional Poisson noise. $\sigma$ is expressed as percentage of the original sinogram mean value.
           The caption of each map specifies what filter and phantom were used 
           for the reconstructions.}
  \label{exp-down-noise-psnr-cubic}
\end{figure}
\hspace{-0.6cm}At visual inspection, the reconstructions with CFBP show better quality and details can be more easily identified.
\newline
CFBP has also been compared to 1D cubic spline interpolation along the view direction (upsampling factor of 2) followed by FBP (IFBP).
Simple 2D interpolation schemes used to double the number of views of a sinogram can yield visible artifacts on the FBP reconstruction
and are, therefore, not considered here for comparison with HLSF. 
The algorithms are tested, first, on noiseless, then, on noisy datasets.
Figure \ref{exp-down-psnr-cubic}(a) shows that CFBP outperforms IFBP for any value of SF, when using the Ram-Lak filter, whereas 
differences vanish, when using the Hanning filter and SF$\,\gtrsim 0.30$, as illustrated by Fig. \ref{exp-down-psnr-cubic}(b).  
Analogously to the results of Fig. \ref{exp-down-noise-psnr}, the stronger the action of the filter, the smaller the SF interval where
CFBP can outperform IFBP for the reconstruction of noiseless undersampled datasets.
The differential maps in Fig. \ref{exp-down-noise-psnr-cubic} are everywhere positive (positive values correspond to 
$\text{PSNR}_{\text{CFBP}}(\sigma,\text{SF}) > \text{PSNR}_{\text{IFBP}}(\sigma,\text{SF})$) and 
show that CFBP yields a better reconstruction quality than IFBP, especially 
for $\sigma>1.73$ and SF$\, < 0.33$.
Very similar results have also been obtained for PH-1, PH-2 and PH-3.
\newline
Other experiments (not shown here) have indicated that no additional improvement can be obtained from either applying HLSF multiple times
sequentially or using HLSF to triple or quadruple at once the number of views.


\section{Discussion and conclusion}
This work presents a fast procedure to improve analytical tomographic reconstructions
of undersampled datasets in parallel beam geometry. The proposed method is a filter working
in the Radon domain and based on the 
Helgason-Ludwig consistency conditions. It doubles the number of views of a sinogram
homogenously sampled in $[0,\pi)$,
by extrapolating projections at intermediate angular positions.
\newline
This sinogram filter, abbreviated HLSF, is a non-iterative, parameterless procedure, that
can be efficiently implemented with FFTs and only marginally impacts the total computational cost for analytical reconstructions.
\newline
Experiments, performed on data of different structural complexity, have shown that FBP reconstructions
of sinograms pre-processed with the presented HLSF are charaterized by a higher PSNR compared to FBP reconstructions of standard sinograms. 
HLSF improves the reconstruction quality for both noiseless (generally, for SF ${\scriptstyle\lesssim}$ 0.31) and noisy undersampled
(especially, for SF ${\scriptstyle\lesssim}$ 0.33 and $\sigma \: {\scriptstyle\gtrsim} \: 1.73\%$) sinograms.
Moreover, HLSF outperforms 1D cubic spline intepolation along the view direction: improvements in the reconstruction accuracy
are substantial when dealing with noisy undersampled datasets (especially, for SF ${\scriptstyle\lesssim}$ 0.33 and $\sigma \: {\scriptstyle\gtrsim} \: 1.73\%$).

\nocite{*}
\bibliographystyle{fundam}
\bibliography{hlsf_fund_inf}


\end{document}